\documentclass{phbauth}
\usepackage{graphicx}

\begin{document}
\begin{frontmatter}
\title{Electrical transport properties of ultrathin disordered films}

\author{G.Sambandamurthy\thanksref{thank1}},
\author{K. DasGupta} and
\author{N. Chandrasekhar}

\address{Department of Physics, Indian Institute of Science,
Bangalore 560 012, INDIA}
\thanks[thank1]{Corresponding author Email:samband@physics.iisc.ernet.in }

\begin{abstract}
We report an experimental study of quench condensed ($2K\le T \le 15K$) 
disordered ultrathin films of {\rm Bi} where localisation effects and 
superconductivity compete. Experiments are done with different substrates
and/or different underlayers. Quasi-free standing films of {\rm Bi}, prepared
by quenching {\rm Bi} vapours onto solid {\rm Xe}, are also studied. The results 
show a dependence of the transport properties both on the dielectric 
constant of the substrate/underlayer as well as the temperature of quench
condensation. RHEED studies indicate that quantum size effects are important in these systems. In this paper, we try to correlate the structure of the films to the transport properties obtained.
\end{abstract}

\end{frontmatter}

\section{Introduction}
The transport properties of 2-D disordered materials have been investigated during the last few decades. One means of varying the disorder in the system is by quench condensing metal flux onto cryogenically cooled substrates (typically $\le 20K$), a technique pioneered
by Buckel and Hilsch \cite{BH}. In this paper, we report studies on ultra thin {\rm Bi} films quench condensed on amorphous quartz substrates with and without {\rm Ge} underlayers. A few monolayers of
{\rm Ge} is thought to improve the wetting properties of the films \cite{HAV}. The temperature of the substrate $T_s$) at which the material is condensed is also varied and the results are presented.
\section{Experiments and Results}
The experiments were done {\em in-situ} in a custom-designed UHV cryostat. The cryostat is pumped by a turbo molecular pump backed by an oil-free diaphragm pump. Base pressure of $\sim$ 
5 $\times$ 10$^{-10}$ Torr can be attained in the system. The substrate temperature can be maintained down to 2K by pumping on the liquid Helium bath. The material ({\rm Bi}) is evaporated from a MBE-type Knudsen cell. A mechanical shutter is used to control the flux. The substrate is pre-deposited with platinum pads
($\sim 50\rm \AA$ thick)for electrical contacts. {\rm Ge} underlayer of 10$\rm \AA$ thickness (if used) is deposited on the substrate before loading it in the cryostat. 
The evolution of sheet resistance ($R_s$) vs temperature is studied. Electrical measurements are done with  standard 4-probe d.c. method.
\begin{figure}[btp]
\begin{center}\leavevmode
\includegraphics[width=1.0\linewidth, height=0.5\linewidth]{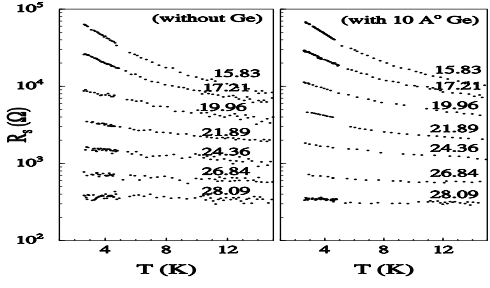}
\caption{ Evolution of $R_s$ vs $T$ for films deposited at $T_s$ = 4.2K. The numbers on the right hand side indicates film thickness in $\rm \AA$}
\label{figurename}\end{center}\end{figure}
Figure 1. shows the evolution of $R_s$ vs $T$ for {\rm Bi} films deposited at $T_s$ = 4.2K. The films grown on 10$\rm \AA$ {\rm Ge} underlayer have higher $R_s$ than corresponding films grown on bare quartz substrates.

\begin{figure}[btp]
\begin{center}\leavevmode
\includegraphics[width=1.0\linewidth, height=0.5\linewidth]{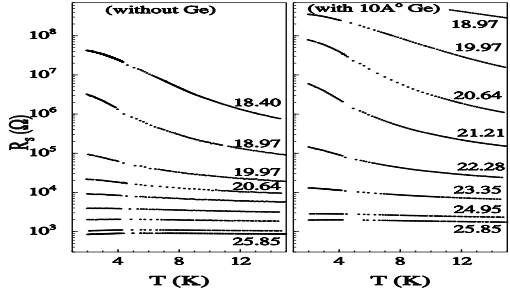}
\caption{ Evolution of $R_s$ vs $T$ for films deposited at $T_s$ = 15K}
\label{figurename}\end{center}\end{figure}
Figure 2. shows the evolution of $R_s$ vs $T$ for films deposited at $T_s$ = 15K. Here the above mentioned effect is even more pronounced.
The inference from these plots are: 
(a)when $T_s$ is higher, the atoms can move through greater distances and rearrange themselves because of higher thermal energy and so it facilitates the formation of bigger clusters. This means less effective coverage of the substrate and so $R_s$ is more, since tunneling would be the dominant conduction mechanism. 
(b)The {\rm Ge} underlayer which is thought to improve the wetting of the films and hence assist
homogenous film growth seems to be doing the opposite. One would, in general, expect the presence of an underlayer to reduce $R_s$ \cite{STR}. But a thin layer of {\rm Ge} offers a number of
dangling bonds for the condensing Bi atoms. Even though the presence of a {\rm Ge} layer provides more
nucleation sites, the {\rm Bi} atoms may attach to dangling bond sites, increasing the atomic scale disorder. Films on bare substrates have disorder on meso or macroscopic scales in contrast to films on {\rm Ge}. The higher atomic scale disorder enhances the possibility of electron localisation and so films have higher $R_s$ \cite{ADK} and the difference of conductance is of the order $\mathrm{e^2/2\pi^2\hbar}$. 
The difference in the behaviour of {\rm Bi} films compared to the previously published results may be attributed to the fact that the {\rm Ge} is deposited at room temperature in our case.
\begin{figure}[btp]
\begin{center}\leavevmode
\includegraphics[width=1.0\linewidth, height=0.5\linewidth]{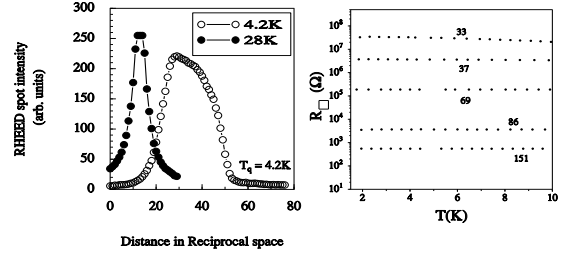}
\caption{a. RHEED horizontal intensity profile of the specularly reflected spot from {\rm Bi}. 
b. Evolution of $R_s$ vs $T$ for films deposited on solid xenon. $T_s$ = 4.2K. {\rm Xe} was quench-condensed on bare quartz.}
\label{figurename}\end{center}\end{figure}
Fig.3.a. indicate that the films consist of clusters that are 25-100 
$\rm \AA$ in diameter.  Films have also been quench-condensed on solid {\rm Xe} underlayers (Fig.3.b.), to study the properties without any interference from the substrate/underlayer. 

\begin{ack}
The work is supported by DST, Government of India. KDG thanks CSIR for the research fellowship.
\end{ack}

\end{document}